\documentclass[]{interact}

\usepackage{epstopdf}

\usepackage{subcaption}
\usepackage{setspace}
\usepackage{xcolor}
\usepackage[round]{natbib}
\usepackage{graphicx}
\usepackage{enumerate}
\usepackage{mathrsfs}
\usepackage{amsthm, amsmath, amssymb,xspace,graphics}
\usepackage{algpseudocode}
\usepackage{algorithm}
\usepackage{animate}
\usepackage{media9}
\usepackage{graphicx}
\usepackage{mwe}
\usepackage{capt-of}

\usepackage[round]{natbib}
\bibpunct[, ]{[}{]}{,}{n}{,}{,}
\makeatletter
\def\NAT@def@citea{\def\@citea{\NAT@separator}}
\makeatother

\theoremstyle{plain}

\theoremstyle{definition}

\theoremstyle{remark}

\newcommand\myeq{\mathrel{\stackrel{\makebox[0pt]{\mbox{\normalfont\tiny ind}}}{\sim}}}
\newcommand{\pro}{_{\text{\tiny prop}}}
\newcommand{\obs}{_{\text{\tiny obs}}}
\newcommand{\init}{_{\text{\tiny init}}}

\newcommand{\FM}{\text{FM}}
\newtheorem{prop}{Proposition}
\usepackage{xcolor}
\begin{document}


\title{Approximate Bayesian Computation for Finite Mixture Models}

\author{
\name{Umberto Simola\textsuperscript{a}\thanks{CONTACT A.~N. Author. Email: umberto.simola@helsinki.fi}, Jessi Cisewski--Kehe\textsuperscript{b} and Robert L. Wolpert\textsuperscript{c}}
\affil{\textsuperscript{a}Department of Mathematics and Statistics, University of Helsinki, Finland; \textsuperscript{b}Department of Statistics, Yale University, New Haven, USA; \textsuperscript{c}Department of Statistical Science, Duke University, Durham, USA}
}

\maketitle

\begin{abstract}
Finite mixture models are used in statistics and other disciplines, but inference for mixture models is challenging due, in part, to the multimodality of the likelihood function and the so-called label switching problem. We propose extensions of the  Approximate Bayesian Computation--Population Monte Carlo (ABC--PMC) algorithm as an alternative framework for inference on finite mixture models. There are several decisions to make when implementing an ABC--PMC algorithm for finite mixture models, including the selection of the kernels used for moving the particles through the iterations, how to address the label switching problem, and the choice of informative summary statistics. Examples are presented to demonstrate the performance of the proposed ABC--PMC algorithm for mixture modeling.  The performance of the proposed method is evaluated in a simulation study and for the popular recessional velocity galaxy data.
\end{abstract}


\section{Introduction} \label{sec.intro}

Mixture models have been used in statistics since the late $1800$s when Karl Pearson introduced them in an analysis of crab morphometry \citep{Pearson1893, Pearson1895}. Subsequently mixture modeling has grown in popularity in the statistical community as a powerful framework for modeling clustered data; McLachlan and Peel \cite{McLachlan2000} provides a general overview of mixture modeling, and
Marin et al. \cite{Marin2005} or Schnatter \cite{Schnatter2006} provide a Bayesian perspective. 
In recent decades mixture models have become routinely applied in various applications \citep{Lancaster1996, Richardson1997, Roeder1997, Henderson2003, Gianola2006, Lin2012, Wu2015}. 
One reason for the general success of mixture models is the ability to specify the number of possibly different component distributions, allowing for flexibility in describing complex systems \citep{Marin2005}.

The general definition for a finite mixture model with fixed integer $K >1$ components is:
\begin{equation}
\sum_{i=1}^{K}f_{i} \cdot p_{i}(y \mid \theta_{i}),
\label{eq:mixture}
\end{equation}
with mixture weights $f_{i} \in (0,1)$ such that $\sum_{i=1}^{K}f_{i}=1$ and $p_{i}(y \mid \theta_{i})$ are the component distributions of the mixture, often parametrically specified with a vector of the unknown parameters, $\theta_{i}$.

Finite mixture models present computational and methodological challenges due, at least in part, to the complex and possibly multimodal likelihood function along with the invariance under permutation of the component indices. The Expectation--Maximization (EM) algorithm \citep{Dempster1977} provides a method for numerically obtaining the maximum likelihood estimates, although the possible \textit{multimodality} of the likelihood function makes finding the global maximum challenging \citep{Marin2005}. Extensions of the EM algorithm have been proposed for improving its speed of convergence and avoiding local optima \citep{Naim2012, Miyahara2017}. 

\begin{sloppypar}
Bayesian methodology and applications for mixture modeling have developed in the last decades \citep{Robert2001,Casella2002, Marin2005,ShabramFord2014,Ramses2015}. Bayesian inference for mixture models often relies on Markov Chain Monte--Carlo (MCMC) which can lead to the so-called \textit{label switching problem} \citep{Diebolt1994, Stephens2000, JasraStephensHolmes2007, Rodriguez2012}, because the likelihood function is invariant to the re-labeling of the mixture components. Additionally the resulting posterior distribution is multimodal and asymmetric, which makes summarizing the posterior distribution using common statistics such as the posterior mean or the Highest Posterior Density (HPD) interval unhelpful \citep{Stephens2000,Stoneking2014,Ramses2015}.
\end{sloppypar}

A different framework for inference that can be considered for addressing the issues related to the use of mixture models is Approximate Bayesian Computation (ABC), whose basic algorithm was first presented conceptually by Rubin \cite{Rubin1984} and computationally by Tavar\'e et al. \cite{TavareEtAl1997} and Pritchard et al. \cite{PitchardEtAl1999}. ABC is often used in situations where the likelihood function is complex or not available, but simulation of data through a forward model (\FM) is possible. The FM is a simulation model that can take parameters as inputs and returns as output simulated data. With mixture models, though the likelihood function can be available, working with the likelihood is not always straightforward.  
Though ABC has its own challenges, we propose an ABC algorithm for estimating the posterior distributions of the parameters providing an alternative to the MCMC algorithms.

Assuming $\theta$ is the possibly multidimensional inferential target, the basic ABC algorithm consists of the following four steps \citep{TavareEtAl1997,PitchardEtAl1999}: (i) sample from the prior distribution, $\theta\pro \sim \pi(\theta)$; (ii) produce a generated sample of the data by using $\theta\pro$ in the \FM, $y\pro \sim \FM(Y \mid \theta\pro)$; (iii) compare the real data, $y\obs$, with the generated sample, $y\pro$, using a distance function, $\rho(\cdot, \cdot)$, letting $d = \rho(s(y\obs), s(y\pro))$ where $s(\cdot)$ is some (possibly multidimensional) summary statistic of the data; (iv) if the distance, $d$, is smaller than a fixed tolerance $\epsilon$ then $\theta\pro$ is retained, otherwise it is discarded.  Repeat until the desired particle sample size $N$ is achieved.

In order to improve the computational efficiency of the basic ABC algorithm, there have been many developments  \citep{SissonEtAl2007,JoyceMarjoram2008,BeaumontEtAl2009,CsilleryEtAl2010,DelMoralEtAl2011,koutroumpas2016bayesian, MarinEtAl2012,RatmannEtAl2013, bonassi2015sequential, toni2009approximate}. In this work, we extend a sequential version of ABC, the ABC--PMC of Beaumont et al. \cite{BeaumontEtAl2009}, which is introduced in the next Section. Interestingly, the original PMC sampler \citep{CappeEtAl2004} was introduced for solving problems related to the multimodality of the likelihood function for settings in which mixture models were used.  This suggests that the ABC--PMC version of the ABC algorithm may be more suitable for mixture models compared to other available algorithms, such as the Sequential Monte Carlo version (ABC--SMC) by \cite{DelMoralEtAl2012}. ABC may be considered preferable over MCMC if the likelihood function is intractable. However, in the finite mixture model setting, the likelihood function is tractable, but there are computational challenges due to the multimodality of the likelihood function as well as the label switching problem that can lead to issues for MCMC \citep{Ramses2015, CappeEtAl2004}.

In any ABC analysis the definition of both suitable summary statistics, $s(\cdot)$, and a distance function, $\rho$,  for comparing the real data  $y\obs$ to the generated sample $y\pro$ is needed and is crucial for getting useful inference results \citep{BeaumontEtAl2002}. The definition of summary statistics is necessary because ABC suffers of the \textit{curse of dimensionality} and using the entire dataset is not computationally feasible \citep{Blum2010a,Blum2010b,BlumEtAl2013, Prangle2015}. 
In the proposed method, we generally use the Hellinger distance for comparing the real data, $y\obs$, to the generated sample, $y\pro$. 
While the proposed method is valid for $y\obs \in \mathbb{R}^{\text{dim}}$, for illustration purposes, we focus most of the discussion on the one dimensional case where $\text{dim}=1$.

The paper is organized as follow. Section \ref{sec.methods} begins with background on the ABC--PMC algorithm by Beaumont et al. \cite{BeaumontEtAl2009}, and then details our proposed extensions for the finite mixture model setting.
Sections \ref{sec.example} and \ref{sec.galaxy} are dedicated, respectively, to a simulation study and a popular real data example using recessional velocities of galaxy data. 
Concluding remarks are presented in Section \ref{sec.conclusions}.

 \section{Methodology} \label{sec.methods}
 
Sequential versions of the basic ABC algorithm have been proposed to improve the computational efficiency of the algorithm. This is typically accomplished by considering a sequence of decreasing tolerances, $\epsilon_{1:T}= (\epsilon_1, \ldots, \epsilon_T)$ such that $\epsilon_1 \geq \epsilon_2 \geq \cdots \geq \epsilon_T$, where $T$ is the total number of iterations, or time steps, of the algorithm. 
Rather than sampling from the prior at each iteration after the initial step, improved proposals are drawn from the ABC posterior sample of the previous step. 
The  ABC--PMC algorithm of Beaumont et al. \cite{BeaumontEtAl2009} relies on ideas of importance sampling  in order to improve the efficiency of the algorithm by constructing a series of intermediate proposal distributions using the sequential ABC posteriors. 
The first iteration of the ABC--PMC algorithm uses tolerance $\epsilon_1$ and draws proposals from the specified prior distributions; from the second iteration through the $T^{\text{th}}$, proposals are selected from the previous iteration's ABC posterior and then moved according to some perturbation kernel (e.g., a Gaussian kernel).  
Since the proposals are drawn from the previous iteration's ABC posterior and moved according to a kernel, rather than being sampled directly from the prior distributions, importance weights are used. The derivation of the importance weight for particle $J = 1, \ldots, N$ in iteration $t$, where $N$ is the desired particle sample size, is discussed in Beaumont et al. \cite{BeaumontEtAl2009} and leads to the following importance weights:
\begin{equation}
W_t^{(J)} \propto \pi(\theta_{t}^{(J)})\Big/\sum_{L = 1}^N W_{t-1}^{(L)} \phi(\theta_{t}^{(J)} \mid \theta_{t-1}^{(L)}, \tau_{\theta,t-1}^2),
\end{equation}
where $\pi(\cdot)$ is the prior distribution, $\phi(\cdot \mid a, b)$ is a Gaussian perturbation kernel with mean $a$ and variance $b$, where  $\tau_{\theta,t-1}^2$ is twice the (weighted) sample variance of the particles for $\theta$ at iteration $t-1$ (as recommended \citep{BeaumontEtAl2009}).  
The importance weights are then standardized so that $\sum_{J = 1}^N W_t^{(J)} = 1$.
Other perturbation kernels can be used and the importance weights would need to be updated accordingly. 
Further details on the original ABC--PMC algorithm can be found in Beaumont et al. \cite{BeaumontEtAl2009}, though there are
several elements that require additional discussion.  In particular, an implementation of ABC--PMC requires the user to select the tolerance sequence $\epsilon_{1:T}$, the perturbation kernels, the summary statistic(s), and the distance function(s).

First, to select the sequence $\epsilon_{1:T}$ we implement an adaptive approach based on the accepted distances from the previous time step. That is, $\epsilon_t$ is set to a particular quantile of the accepted distances from time step $t-1$ \citep{DelMoralEtAl2012,WeyantEtAl2013,Lenormand2013,IshidaEtAl2015, Cisewski-Kehe:2019aa, simola2019machine}.
Selecting too high of a quantile at which to shrink the tolerance could result in a decrease in computational efficiency because \emph{more iterations} would be needed to shrink the tolerance to a small enough value.
On the other hand, selecting a quantile that is too low can also contribute to computational inefficiencies because \emph{more draws} from the proposal are needed within each iteration to find datasets that achieve the small tolerance. 
The rule used to shrink the tolerance is important as imprudent selection can lead to the ABC--PMC algorithm getting stuck in local modes \citep{Silk2013,SimolaEtAl2019}. 
This latter potential issue is particularly crucial when working with mixture models, since the multimodality of the likelihood function is a well known problem that needs to be addressed in order to produce reliable statistical inference results.   

The decreasing tolerance sequence's impact on the achievement of the global maximum is addressed in two ways in the proposed algorithm.
First, we initialize the tolerance sequence by oversampling in the first iteration of the ABC--PMC algorithm \citep{Cisewski-Kehe:2019aa}. 
Let $N$ be the desired number of particles used to approximate the posterior distribution, then the initial tolerance, $\epsilon_{1}$, can be adaptively selected by sampling $N\init = lN$ draws from the prior, for some $l \in \mathbb Z^+$.  
Then the $N$ particles of the $lN$ total particles with the smallest distances are retained, and $\epsilon_1 = \max \left(d_1^{(1*)}, \ldots, d_1^{(N*)}\right)$ where $d_1^{(1*)}, \ldots, d_1^{(N*)}$ are the $N$ smallest distances of the $lN$ particles sampled.
For all examples considered in this work, we use $l=5$. 
Using this adaptive initialization, the parameter space is better sampled than if we only considered $N$ draws from the prior distribution. 
Secondly, we use lower quantiles for the first several iterations of the ABC--PMC algorithm per the suggestion in Silk et al. \cite{Silk2013} and Simola et al. \cite{SimolaEtAl2019}, which can help the algorithm avoid local modes.

One of the challenges in using ABC--PMC for mixture models is related to selecting an appropriate perturbation kernel for the mixture weights because the individual weights must be between 0 and 1, and the weights must sum to 1.  Hence, the usual Gaussian perturbation kernel is not a viable option because this kernel can lead to proposed mixture weights that are not within the noted constraints. 
An additional challenge in using ABC--PMC for mixture models is due to the label switching problem, and for reasons discussed below, this has to be addressed at the end of each iteration.  Finally, the selection of an appropriate and informative summary statistic for multimodal data requires a summary that can capture the overall shape of the distribution of the data.

For these reasons the original version of the ABC--PMC algorithm is not directly able to accommodate mixture models. Our proposed extensions are discussed in more detail later, but can be summarized as follows. In Algorithm \ref{alg1}, we adapt the ABC--PMC algorithm to work with finite Gaussian mixture models. In particular, we propose a procedure for resampling the mixture weights in a way that capitalizes on distributional properties of the prior distribution for the mixture weights (i.e., the Dirichlet distribution), which is presented in Algorithm \ref{alg:dr}. The mixture weights could be resampled using other methods, such as resampling the particles according to a Gaussian perturbation kernel, but the resampled weights would need to be adjusted in order to preserve properties of weights such as being between or equal to 0 and 1 and summing to 1. Additionally, in order to address the label switching problem, we designed a procedure that is executed at the end of each iteration of the ABC--PMC algorithm, which is presented in Algorithm \ref{alg:ls}. An overview of different available algorithms designed for addressing the label switching problem and more details on Algorithm \ref{alg:ls} are presented in Section \ref{subsec.ls}.  If not properly addressed, the label switching problem can hinder the algorithm's capability of finding the true posterior distribution.

\subsection{Finite Gaussian Mixture Models}
A common choice for the $p_{i}(\cdot \mid \cdot)$ of Eq.~\eqref{eq:mixture} is the Gaussian distribution. This particular class of models, called Gaussian Mixture Models (GMMs), is  flexible and widely used in various applications (e.g., \citep{Gianola2006, Lin2012,Stoneking2014}). Maintaining the notation of Eq.~\eqref{eq:mixture}, a GMM is defined as:
\begin{equation}
\sum_{i=1}^{K}f_{i} \cdot \phi(y \mid \theta_{i}),
\label{eq:GaussiaKixture}
\end{equation}
where $\phi$ is the density function of the Gaussian distribution, $\theta_{i} = (\mu_{i}, \sigma_{i}^{2})$ is the vector of unknown means and variances for each of the $K$ groups with the unknown parameters  $\mu = (\mu_{1}, \dots, \mu_{K}) \in \mathbb{R}^{K}$, $\sigma^{2} = (\sigma_{1}^{2}, \dots, \sigma_{K}^{2}) \in \mathbb{R}_{+}^{K}$, and the mixture weights, $f =(f_{1}, \dots, f_{K})$. Note that the $f =(f_{1}, \dots, f_{K})$ lie in the unit simplex, $\Delta^{K-1}\equiv\{x\in\mathbb{R}^K_+:~ \sum_j x_j=1\}$ inside the unit cube $[0,1]^K$. 

A common choice of the prior distribution for $f=(f_{1}, \dots, f_{K})$ is the Dirichlet distribution of order $K$ with hyperparameter $\delta=(\delta_1,\dots \delta_K)$, where often $\delta \equiv (1, \dots, 1)$, as proposed by Wasserman \cite{Wasserman2000}. Another common choice has been proposed by Rousseau and Mengersen \cite{Rousseau2011}, who defined the hyperparameter $\delta \equiv (1/2, \dots, 1/2)$; in this way the prior is marginally a Jeffreys prior distribution. The priors for the mean and the variance of the GMM from Eq.~\eqref{eq:GaussiaKixture} can be defined as follows:
\begin{equation}
\mu_{i} \sim \phi(\mu \mid \xi,\kappa), \qquad
\sigma_{i}^{2} \sim \text{\text{Inv--Gamma}}(\alpha,\beta),
\label{eq:prior.sigma}
\end{equation}
with mean $\xi$, variance $\kappa$, and shape parameter $\alpha$ and rate parameter $\beta$. 
There are several methods for selecting the hyperparameters, $\eta = (\xi, \kappa, \alpha, \beta)$, such as the Empirical Bayes approach \citep{Casella2000} and the `weakly informative principle' \citep{Richardson1997}. Both of these options are considered in examples later so as to be consistent with the original analysis of the example.

Our proposed algorithm is specified in Algorithm~\ref{alg1} for the model defined in Eq.~\eqref{eq:GaussiaKixture}, and the details of important steps are presented next.

\begin{algorithm}
\caption{ABC--PMC for Finite Gaussian Mixture Model}
\begin{algorithmic}  
\State Select the number of components $K$
\State Select the desired number of particles $N$
\State Select the desired number of particles coming from the prior $N\init$, $N\init>N$, for $t=1$
\If{$t = 1$}
\For{$J = 1, \ldots, N\init$}
 	\State Propose $\mu^{(J)}_1 = \{\mu_{1,1}, \ldots, \mu_{{1,K}}\}^{(J)}$ by drawing from prior $\mu_{k}^* \sim \pi(\mu)$, $k = 1, \ldots, K$
	\State Propose $\sigma^{2^{(J)}}_1 = \{\sigma^{2}_{1,1}, \ldots, \sigma^{2}_{{1,K}}\}^{(J)}$ by drawing from prior $\sigma_{k}^{2^{*}} \sim \pi(\sigma^2)$, $k = 1, \ldots, K$
	\State Propose $f^{(J)}_1 = \{f_{1,1}, \ldots, f_{{1,K}}\}^{(J)}$ by drawing from prior $f_{k}^* \sim \pi(f)$, $k = 1, \ldots, K$
 	\State Generate  $y\pro$ from $\sum_{i=1}^{K}f^{(J)}_{1,i} \cdot \phi(y \mid \mu^{(J)}_{1,i},\sigma^{2^{(J)}}_{1,i})$
 	\State Calculate distance $d_1^{(J)} = \rho\left\{y\obs, y\pro\right\}$
 \EndFor	
	\State Put $d_1$ in increasing order and set $\epsilon_{1}=d_{1}^{(N)}$, where $(N)$ is the $N^{\text{th}}$ smallest distance
	\State Retain corresponding elements $\mu^{(1:N)}_1, \sigma^{2^{(1:N)}}_1, f^{(1:N)}_1$, the proposed values corresponding to the $N$ smallest distances
 	\State Set weight $W_1^{(J)} = N^{-1}$
	\State Address the label switching problem (see Algorithm \ref{alg:ls} in \S \ref{subsec.ls})
 \ElsIf{$2 \leq t \leq T$}
 \For{$J = 1, \ldots, N$}
 \State Set $\epsilon_t = q^{\text{th}}$ quantile of $\left\{d_{t-1}^{(J)}\right\}_{J = 1}^N$
 \State Set $d_t^{(J)} = \epsilon_t + 1$
 \While{$d_t^{(J)}> \epsilon_t$}
	\State Select $\left\{f_t^*,\mu_t^*,\sigma^{2^{*}}_t\right\}$ from $\left\{f_{t-1}^{(J)},\mu_{t-1}^{(J)},\sigma^{2^{(J)}}_{t-1}\right\}_{J = 1}^N$ with probabilities $\left\{W_{t-1}^{(J)}/\sum_{M = 1}^NW_{t-1}^{(M)}\right\}_{J = 1}^N$ 
	\State Propose $f_t^{(J)}$ according to the Dirichlet resampling functions (see Algorithm \ref{alg:dr} in \S \ref{subsec.dirichlet.res})
	\State Propose $\mu_t^{(J)} \sim \phi_{\text{K}}(\mu \mid \mu_t^* ,  \tau^{2}_{\mu, t-1})$, where $\phi_{\text{K}}$ is a multivariate Gaussian density.
	\State Propose $\sigma_t^{2^{(J)}} \sim \phi_{\text{K}}(\sigma^2 \mid \sigma_t^{2^{*}} ,  \tau^{2}_{\sigma^2, t-1} \text{ with } \sigma^2 > 0)$ 
	\State Generate  $y\pro$ from $\sum_{i=1}^{K}f^{(J)}_{t,i} \cdot \phi(y \mid \mu^{(J)}_{t,i},\sigma^{2^{(J)}}_{t,i})$ 
 	\State Calculate distance $d_t^{(J)} = \rho\left\{y\obs, y\pro\right\}$
	\State Address the label switching problem (see Algorithm \ref{alg:ls} in \S \ref{subsec.ls})
 \EndWhile
	\State Set weight $W_t^{(J)} \propto \prod_{i=1}^K \pi(\mu_{t,i}^{(J)}) \pi(\sigma^{2^{(J)}}_{t,i})/\sum_{M = 1}^N W_{t-1}^{(M)} \phi_{\text{K}}(\mu_{t}^{(J)} \mid \mu_{t-1}^{(M)},\tau_{\mu, t-1}^2) \phi_{\text{K}}(\sigma_{t}^{(J)} \mid \sigma_{t-1}^{(M)}, \tau_{\sigma, t-1}^2 \text{ with } \sigma_{t}^{(J)} > 0 )$
 \EndFor
 \EndIf
\end{algorithmic} \label{alg1}
\end{algorithm}

\subsection{Perturbation kernel functions} \label{subsec.dirichlet.res}
One of the advantages of ABC--PMC over the basic ABC algorithm is that, starting from the second iteration, rather than drawing proposals from the prior distribution, proposed particles are drawn from the previous step's ABC posterior according to their importance weights. 
Then, instead of using the actual proposed value that was drawn, it is perturbed according to some kernel. Constructing computationally efficient perturbation kernels is crucial to improve the computational performance of the ABC algorithm  \citep{filippi2013optimality}.
There are a number of kernel functions, $K(\cdot \mid \cdot)$, available for perturbing the proposed particles. Beaumont et al. \cite{BeaumontEtAl2009} suggest a Gaussian kernel centered on the selected particle from the previous iteration and a variance equal to twice the empirical variance of the previous iteration's ABC posterior. This is a reasonable choice if there are no constraints on the support of the parameter of interest. 
For the means and the variances of the mixture model, a multivariate Gaussian distribution and a multivariate Gaussian distribution truncated to $\mathbb{R}_{+}$  were used, respectively. Both perturbation kernels are assigned variances per the suggestion from Beaumont et al. \cite{BeaumontEtAl2009} (i.e., variances set to be twice the empirical variances of the previous iteration's ABC posteriors).
When constraints on the parameter support are present, such as for variances or mixture weights, a perturbation kernel should be selected so that it does not propose values outside the parameter's support. 
%
When moving the selected values for the mixture weights, not only is there the constraint that each mixture weight component must be in $[0,1]$, but it is also required that $\sum_{i = 1}^{K} f_{i} = 1$, making the Gaussian kernel inappropriate.

\begin{sloppypar}
In the first iteration of the proposed ABC--PMC algorithm and for $J = 1, \dots, N\init$, the mixture weights $f^{(J)}_1 =  \{f_{1,1}, \ldots, f_{{1,K}}\}^{(J)}$
are directly sampled from the prior distribution, which is a $\text{Dirichlet}(\delta$), where $\delta = (\delta_{1}, \dots, \delta_{K})$. 
For $t>1$ and for $J = 1, \dots, N$, proposals are drawn from the previous iterations's particle system according to their importance weights, as defined in Algorithm~\ref{alg1}. 
After randomly selecting some particle $L \in \{1, \ldots, N\}$ from the $t-1$ iteration's particle system we want to ``jitter'' or move the selected particle's mixture weight, 
$f_t^*$, in manner that preserves some information coming from the selected particle, but not let it be an identical copy, in order to obtain the resampled mixture weights, 
$f_t^{(L)}$. This is carried out using Algorithm \ref{alg:dr} and the mathematical justification can be found in Appendix~\ref{appendix}.

In Algorithm \ref{alg:dr}, for each resampled mixture weight for iteration $t = 2, \ldots, T$ and mixture component $i = 1, \ldots, K$, rely on the random variable denoted $\xi^{*}_{t,i}$ (the particle index $J$ is suppressed for notational convenience in what follows), which is designed to be the sum of two independent random variables, $Z_t f_t^*B_{t,i}  \sim \text{Gamma}(p\delta_{i},1)$ and $\eta_{t,i} \sim \text{Gamma}((1-p)\delta_{i},1)$.
The resampled mixture weight is $f_t =  \xi_{t,i}^*/\xi_{t,+}^*$, with $\xi_{t,+}^*=\sum_{i=1}^{K} \xi_{t,i}^*$,
 so that $f_t \sim \text{Dirichlet}(\delta)$.  
The key benefits of this procedure are that the resampled weight vector, $f_{t}$, retains some information from the $f_t^*$ (i.e., the previous iterations particle that was selected for resampling) through the parameter $p$ (discussed next), but also follows the same distribution as  the prior, $\text{Dirichlet}(\delta)$, so that the importance weights due to this resampling are simply unity.

The parameter $p$ is a fixed real number with range $[0,1]$ that determines how much information to retain from $f_t^*$. The choice of $p$ has an impact on both the allowed variability of the marginal ABC posterior distributions for the mixture weights and the efficiency of the entire procedure. In particular, fixing a $p$ close to $1$ leads to a Dirichlet resampling in which the new set of mixture weights $f_t$ is close to the previous set $f_t^*$ (if $p=1$, then $f_t= f_t^*$). On the other hand a choice for $p$ close to 0 implies that the information coming from $f_t^*$ is weakly considered (for $p=0$ a new set of particles is drawn directly from the prior distribution and no information about $f_t^*$ has been retained). We found $p=0.5$ to be a good choice for balancing efficiency and variability (i.e., it allows for some retention of information of the selected particle, but does not lead to nearly identical copies of it). 

\end{sloppypar}

\begin{algorithm}
\caption{Resampling the mixture weights to obtain $f_t^{(J)} = \{f_{t,1}, \ldots, f_{t,K}\}^{(J)}$}
\begin{algorithmic}  
\State Given $f_t^*$  randomly selected from $\{ f_{t-1}^{(J)}\}_{J = 1}^N$ with probabilities $\left\{W_{t-1}^{(J)}/\sum_{M = 1}^NW_{t-1}^{(M)}\right\}_{J = 1}^N$ 
\State 1. Draw $Z_t \sim \text{Gamma}(\delta_{+},1)$, with $\delta_{+}=\sum_{i=1}^{K} \delta_{i}$ and set $\xi_{t,i} = Z_{t} f_t^*$. 
Then $\xi_{t,i} \myeq \text{Gamma}(\delta_{i},1), i = 1, \ldots, K$.
\State 2.  Select a real number $p \in [0,1]$. (We suggest $p=0.5$.)
\State 3. Draw $B_{t,i} \sim \text{Beta}(p\delta_{i},(1-p)\delta_{i})$ independently for $i = 1, \dots, K$, where $\xi_{t,i}B_{t,i} \myeq \text{Gamma}(p\delta_{i},1)$.
\State 4. Draw $\eta_{t,i} \myeq \text{Gamma}((1-p)\delta_{i}, 1)$ independently.
\State 5. Set $\xi^{*}_{t,i} = Z_t f_t^*B_{t,i} + \eta_{t,i}$ and $f_t^{(J)} = \xi_{t,i}^*/\xi_{t,+}^*$, with $\xi_{t,+}^*=\sum_{i=1}^{K} \xi_{t,i}^*$.  Then $f_t^{(J)} = \{f_{t,1}, \ldots, f_{t,K}\}^{(J)} \sim \text{Dirichlet}(\delta_1, \ldots, \delta_K)$.
\end{algorithmic} \label{alg:dr}
\end{algorithm}

\subsection{Algorithm for addressing the label switching problem} \label{subsec.ls}
\begin{sloppypar}
As noted earlier, a common problem arising in the Bayesian mixture models framework is label switching. When drawing a sample from a posterior (for both MCMC and ABC), the sampled values are not necessarily ordered according to their mixture component assignments because the likelihood is exchangeable. 
For example, suppose a particle 
$\{(f_1^{(J)}, \ldots, f_K^{(J)}), (\mu_1^{(J)}, \ldots, \mu_K^{(J)}), (\sigma^{2^{(J)}}_1, \ldots, \sigma^{2^{(J)}}_K)\}$,  where $J \in \{1, \ldots, N\}$ , is accepted for a $K$ component GMM.  (Note that in this Section, we do not include the algorithm step index, $t$, to simply the notation.)  This particle was selected with a particular ordering of the $1, \ldots, K$ components with $f_i^{(J)}$, $\mu_i^{(J)}$, and $\sigma^{2^{(J)}}_i$ coming from the same mixture component, $i = 1, \ldots, K$; however, a new particle that is accepted will not necessarily follow that same ordering of the $i = 1, \ldots, K$ components.
Somehow the particles have to be ordered in such a way that aligns different realizations of the $i = 1, \ldots, K$ components in order to eliminate the ambiguity. 
\end{sloppypar}

\begin{sloppypar}
Several approaches have been proposed to address the label switching problem and are known as relabeling algorithms.  A first group of relabeling algorithms consists of imposing an artificial identifiability constraint in order to arbitrarily pick a parameter (e.g., the mixture weights) and sort all the parameters for each accepted particle according to that parameter's order \citep{Diebolt1994,Richardson1997}. However, many of the algorithms proposed for addressing the label switching are deterministic (e.g., Stephen's method \citep{Stephens2000} and the pivotal reordering algorithm \citep{Marin2005}). A third class of strategies, called probabilistic relabeling algorithms, uses a probabilistic approach for addressing the label switching problem \citep{Sperrin2010}. A detailed overview of methods for addressing label switching is presented in Papastamoulis \cite{Papastamoulis2016}.
In Section \ref{sec.example.1}, we provide an example that illustrates a problem with the artificial identifiability constraint approach.
Instead, we propose a deterministic strategy for addressing the label switching by selecting a parameter that has at least two well-separated components.  
\end{sloppypar}

Addressing the label switching problem is especially important for the proposed sequential ABC algorithm because each time step's ABC posterior is used as the proposal in the subsequent step of the algorithm so the label switching has to be resolved before using it as a proposal distribution.
Algorithm \ref{alg:ls} outlines the proposed strategy, and is carried out at the end of each iteration. 
The key aspect of Algorithm \ref{alg:ls} is to select a parameter for reordering that has at least two well-separated components. 
To determine this, each set of parameters (e.g., the means, the variances, the mixture weights), is arranged in increasing order. For example, for each particle $J$, $J=1,\ldots,N$, $\mu^{(J)}$ would be ordered so that $\mu_{(1)}^{(J)} \le \mu_{(2)}^{(J)} \le \cdots \le \mu_{(K)}^{(J)}$, with $\mu_{(i)}^{(J)}$ as the $i^{\text{th}}$ order statistic. This is carried out for each set of parameters with analogous notation.

\begin{sloppypar}
The next step is to determine which set of parameters has the best separated components values. We propose first shifting and scaling each set of parameters to be supported within the range $[0,1]$ so that scaling issues are mitigated and the different parameter set values are comparable. 
One option for this adjustment is to use some distribution function, such as a Gaussian distribution with a mean and variance equal to the sample mean and the sample variance of the considered parameter set (e.g., the sample mean for the $\mu$'s is $\bar{\mu}=(NK)^{-1} \sum_{k=1}^{K} \sum_{J=1}^{N} \mu_{(k)}^{(J)}$, and the sample variance for the $\mu$'s is $\text{var}{(\mu)}=(NK)^{-1} \sum_{k=1}^{K} \sum_{J=1}^{N} (\mu_{(k)}^{(J)}-\bar{\mu})^2$). 
The resulting $k^{\text{th}}$ largest standardized value for the mixture mean is $\widetilde{\mu}^{(J)}_{(k)}$. This is carried out for each set of parameters with analogous notation. 
\end{sloppypar}

For each component of each ordered and standardized particle a representative value, such as a mean, is computed (e.g., $\bar{\widetilde{\mu}}_{(k)} = N^{-1}\sum_{J = 1}^N \widetilde{\mu}^{(J)}_{(k)}$ is the representative value of the $k^{\text{th}}$ component of the mean parameter). This is carried out for each set of parameters with analogous notation. 
The pairwise distances (pdist) between the representative values within each parameter set is determined.  
The parameter that has the largest separation between any two of its representative values is selected for the overall ordering of the particle system.
\begin{algorithm}
\caption{Addressing the label switching problem}
\begin{algorithmic}  
\State 1. For each parameter set, obtain the ordered particles $\mu^{(J)}_{(k)}$,  $\sigma^{2^{(J)}}_{(k)}$ and  $f^{(J)}_{(k)}$, $k = 1, \ldots, K$, $J = 1, \ldots, N$
\State 2. Shift and rescale each set of parameters to be supported within the range $[0,1]$, retrieving $\widetilde{\mu}^{(J)}_{(k)} = \Phi(\mu^{(J)}_{(k)} \mid \bar{\mu}, \text{var}(\mu))$, $\widetilde{\sigma}^{2^{(J)}}_{(k)} = \Phi(\sigma^{2^{(J)}}_{(k)} \mid \bar{\sigma}^{2},\text{var}(\sigma^{2}))$ and $\widetilde{f}^{(J)}_{(k)} = \Phi(f^{(J)}_{(k)} \mid\bar{f},\text{var}(f))$, where $\Phi(\cdot \mid a, b)$ is the Gaussian distribution function with mean $a$ and variance $b$.
\State 3. Compute representative values (such as a mean) for each shifted and standardized component,  $\bar{\widetilde{\mu}}_{(k)}$, $\bar{\widetilde{\sigma}}^{2}_{(k)}$ and $\bar{\widetilde{f}}_{(k)}$
\State 4. Compute the pairwise distances (pdist) within each set of representative values,  $\bar{\widetilde{\mu}}_{(k)}$, $\bar{\widetilde{\sigma}}^{2}_{(k)}$ and $\bar{\widetilde{f}}_{(k)}$ 
\State 5. The overall ordering of the particle system is based on the ordering of the parameter set with the largest separation between any two of its representative values
\end{algorithmic} \label{alg:ls}
\end{algorithm}

Variations of Algorithm~\ref{alg:ls} were considered.  
For example, rather than sorting based on the parameter set with largest separation between any two of its representative values, we considered basing the sorting on the parameter with the largest separation between its two \emph{closest} representative values (i.e., the maximum of the minimum separations); however, this sorting did not perform well empirically.  
The issue seemed to be that  parameter with the largest separation between its two closest representative values may actually have all of its components relatively close; after multiple iterations, none of the components were well separated  from the other components.  This lead to iteration after iteration of components that remained a blend of components rather than separating clearly into $K$ components.
Overall, from empirical experiments, the approach outlined in Algorithm~\ref{alg:ls} performed the best and thus is our recommendation.  However, we emphasize that alterations to this procedure may be necessary for mixture models that have additional structure or correlations among the parameters of the component distributions or to deal with multi--dimensional mixture models.

\subsection{Summary statistics} \label{subsec.ss}

Comparing the entire simulated dataset, $y\pro$, with the entire set of observations, $y\obs$, in an ABC procedure is not computationally feasible. For this reason the definition of a lower dimensional summary statistic is necessary. 
For mixture models, due to the multimodality of the data, common summaries such as means or higher order moments (e.g., variance, skewness or kurtosis) do not capture relevant aspects of the distribution which can lead to incorrect ABC posterior distributions. However, an estimate of the density of the data can better account for its key features (e.g., the shape of
each component of the mixture). For the examples presented in Sections \ref{sec.example} and \ref{sec.galaxy}, we suggest using kernel density estimates (KDEs, \citep{Wasserman2006}) of the simulated data, $\hat{f}_{y_{\text{prop}, n}}$, and the observations, $\hat{f}_{y_{\text{obs}, n}}$, to summarize the data, and then the Hellinger distance, $H$, to carry out the comparison. We considered other distances, such as the Kolmogorov--Smirnov (KS) distance \citep{kolmogorov1933sulla,smirnov1948table}, but found the Hellinger distance to perform better than any other tested distance function. The Hellinger distance quantifies the similarity between two density functions, $f$ and $g$, and is defined as:
\begin{equation}
H(f,g) = \left(\int \left(\sqrt{f(y)}-\sqrt{g(y)} \right)^2 dy \right)^{\frac{1}{2}}.
\label{eq:hellinger}
\end{equation}
At each iteration $t$ of the proposed ABC--PMC procedure, a proposed $\theta$ is accepted if $H(\hat{f}_{y_{\text{obs}, n}},\hat{f}_{y_{\text{prop}, n}}) < \epsilon_{t}$, where $\epsilon_{t}$ is the tolerance.

\section{Simulation study} \label{sec.example}

In this Section a simulation study is presented to evaluate the behavior of the proposed ABC--PMC algorithm, Algorithm~\ref{alg1}, using popular GMM examples. 
In particular we are interested in evaluating the success of the procedure with respect to the label switching problem and the reliability of the Hellinger distance as a summary statistic. 
To determine the number of iterations, $T$, a stopping rule was defined based on the Hellinger distance between sequential ABC posteriors; once the sequential Hellinger distance dropped below a threshold of $0.05$ for each of the marginal ABC posteriors, the algorithm was stopped.

\subsection{Mixture Model with equal and unequal group sizes} \label{sec.example.1}
The first example is taken from Mena and Walker \cite{Ramses2015}, which considered a GMM obtained by simulating data coming from $K=2$ groups of equal size which was designed for evaluating performance with respect to the label switching problem.
The 40 observations were simulated as follows:
\begin{equation}
Y_{i=1,\dots,20} \sim \phi(y \mid -20,1),  \hspace{.25cm} Y_{i=21,\dots,40} \sim \phi(y \mid 20,1),
\label{eq:mxtureRamses1}
\end{equation}
where $\phi$ is the Normal density function as defined previously. 
The variance is assumed known for both groups and hence the parameters of interest are the mixture weights, $f = (f_{1}, f_{2})$ (with $f_{2} = 1 - f_{1}$), and the means $\mu = (\mu_{1}, \mu_{2})$. 
The prior distributions selected for this analysis are
$\mu_{i} \sim \phi(\mu \mid 0,100)$, 
and
$f=(f_1, f_{2}) \sim \text{Dirichlet}(1,1)$.
The number of particles is set to $N=5000$ and the quantile used for shrinking the tolerance is $q=0.5$. The algorithm was stopped after $T=20$ iterations, since further reduction of the tolerance did not lead to a notable improvement of the ABC posterior distributions (evaluated by calculating the Hellinger distance between the sequential ABC posterior distributions as noted previously).  

\begin{figure}[!ht]
    \centering
    \begin{subfigure}{6cm}
    \includegraphics[height = 2.5in]{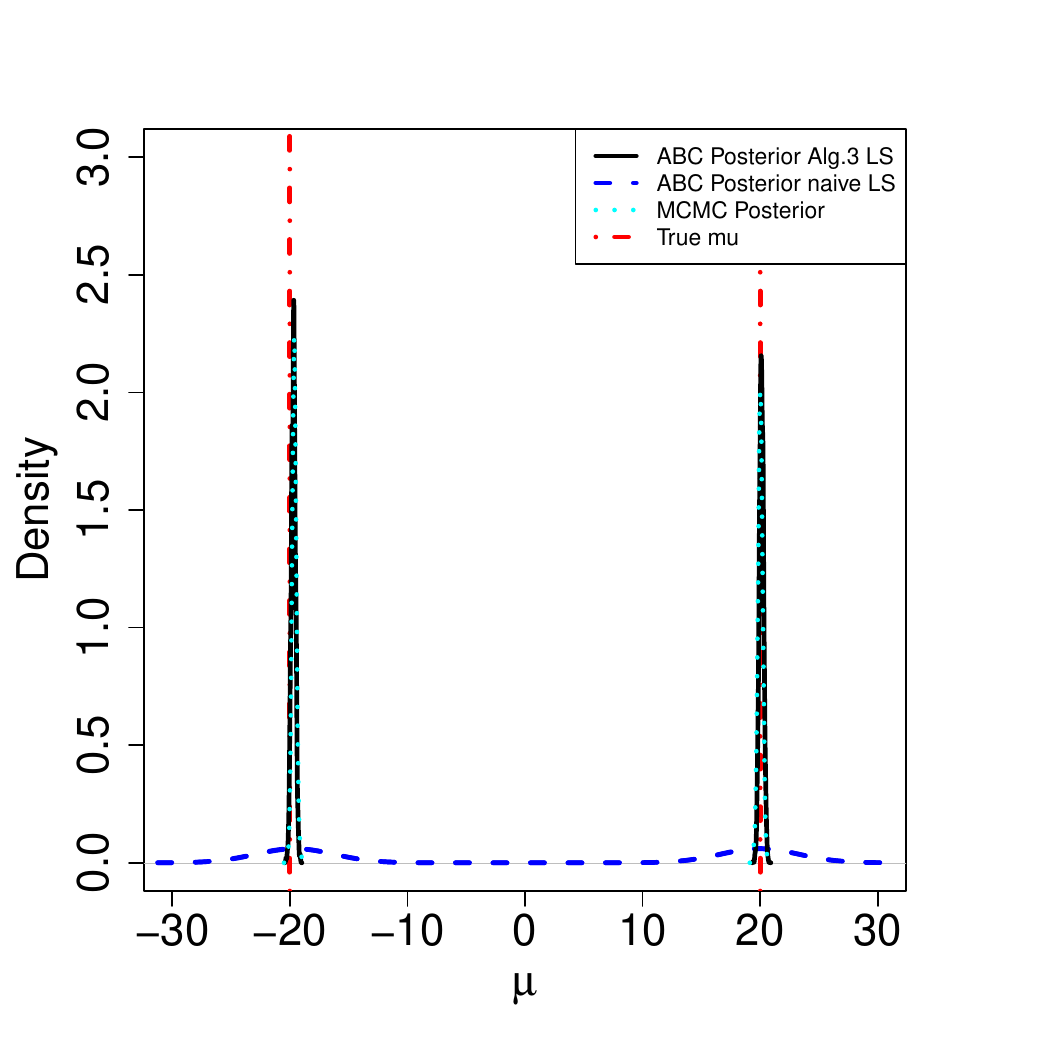}
    \caption{Marginal posteriors for $\mu$}\label{fig:ex1.mean}
    \end{subfigure}
    \qquad
    \begin{subfigure}{6cm}
    \includegraphics[height = 2.5in]{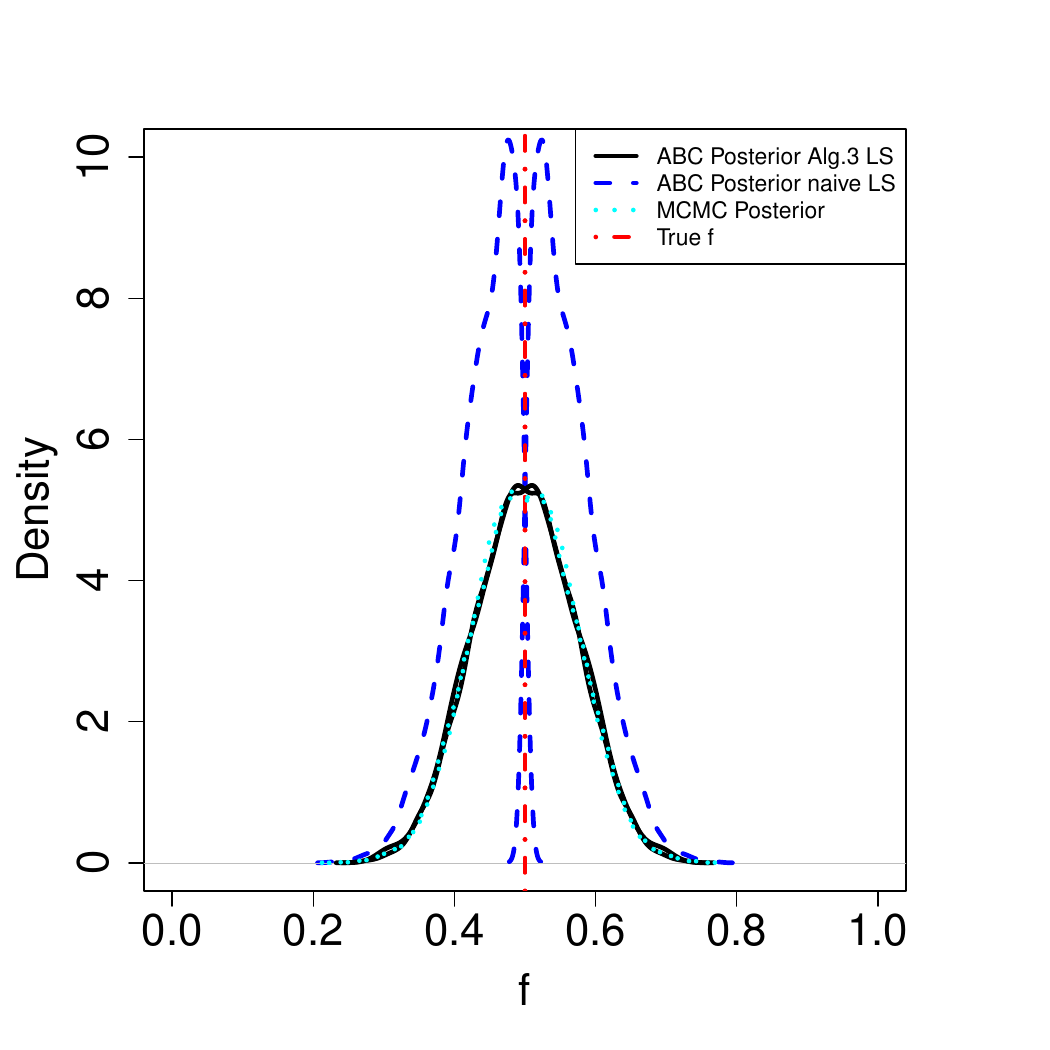}
    \caption{Marginal posteriors for $f$}\label{fig:ex1.f}
    \end{subfigure}
    \caption{Comparison between the ABC and the MCMC marginal posterior distributions for the two-component GMM example from Mena and Walker \cite{Ramses2015}. The true values for the means and the mixture weights are displayed as vertical dot-dashed red lines.
The final ABC posteriors obtained using the label switching (LS) procedure proposed in \S\ref{subsec.ls} are the solid black lines (ABC Posterior Alg.3 LS), and the naive approach that sorts based on the mixture weights are the dashed blue lines (ABC Posterior naive LS). The MCMC posteriors are displayed with dotted cyan lines (MCMC Posterior).
   The number of particles for the ABC analysis and the number of elements kept from the MCMC analysis (after the burn-in) are equal to $5000$. 
   }%
    \label{Fig:example1}%
\end{figure}

Figure \ref{Fig:example1} displays the resulting ABC posteriors (`ABC Posterior Alg. 3 LS' and `ABC Posterior naive LS' are discussed later for illustrating the label switching issue) 
and the corresponding MCMC posteriors, which are used as a benchmark to access the performance of the proposed method. The ABC posteriors closely match the MCMC posteriors. The summary of the results presented in Table~\ref{Table:example1} demonstrate that the ABC posterior distributions are a suitable approximation of the MCMC posteriors. 
The Hellinger distances between the marginal ABC and MCMC posteriors are also displayed in the last column of Table~\ref{Table:example1}; the Hellinger distance between the MCMC and the ABC posterior is $0.032$ for the mixture weights and $0.21$ for the means. 

\begin{center}
\begin{table}
\centering
\begin{tabular}{|c|c|c|c|}
\hline
Parameter (input)           & MCMC (SD)          &   ABC (SD)   & H  \\
\hline
$f_{1} (0.5)$            &    $0.5008 (0.076)$    & $0.5003 (0.076)$ & $0.032$    \\
\hline
$f_{2}  (0.5)$            &     $0.4991(0.076)$   &  $ 0.4996 (0.076) $ &   $0.032$    \\
\hline
$\mu_{1}  (-20)$      &     $-19.72 (0.18)$    &  $-19.70 (0.18)$   & $0.21$      \\
\hline
$\mu_{2} (20)$      &     $20.05 (0.19)$     &  $20.10 (0.19)$   &  $0.21$     \\
\hline
\end{tabular}
\caption{Posterior means (posterior standard deviations) obtained by using the MCMC and the ABC--PMC algorithm for the two-component GMM example from Mena and Walker \cite{Ramses2015}. 
The fourth column indicates the Hellinger distance between the final ABC  and the MCMC posteriors. The number of ABC particles and the number of elements retained from the MCMC chain (after the burn-in) are both equal to $5000$.}
\label{Table:example1}
\end{table}
\end{center}

As noted in Section~\ref{subsec.ls}, the label switching problem has to be carefully addressed when using the ABC--PMC algorithm. For each time step following the initial step, the previous step's ABC posterior is used as the proposal rather than the prior distribution so the procedure for addressing the label switching proposed in Section~\ref{subsec.ls}, Algorithm~\ref{alg:ls}, is used at the end of each time step.
In order to illustrate the consequences of incorrectly addressing the label switching, we ran the proposed ABC algorithm on the example proposed by Mena and Walker \cite{Ramses2015}, except rather than using Algorithm~\ref{alg:ls}, the ordering of the particle system is carried out using the ordering of the mixture weights; the mixture weights are both equal to $0.5$ for this example, making them a poor choice for attempting to separate out the mixture components.  
The resulting ABC posteriors are displayed in Figure~\ref{Fig:example1} (blue dashed lines).  The means, $\mu = (\mu_1, \mu_2)$, of the mixture components are shuffled and not close to the MCMC posterior, while the mixture weights are sorted such a way all the elements of $f_{1}$ are smaller than $0.5$ and all the elements of $f_{2}$ are larger than $0.5$. 

Mena and Walker \cite{Ramses2015} added a third component to the mixture in Eq.~\eqref{eq:mxtureRamses1}, simulating five additional observations from a standard Gaussian distribution, obtaining a three-component GMM with known variances. The ABC--PMC algorithm was run with the same specifications as the first part of the example, but required $T=25$ steps to achieve our stopping rule.

\begin{figure}[!ht]
    \centering
    \begin{subfigure}{6cm}
    \includegraphics[height = 2.5in]{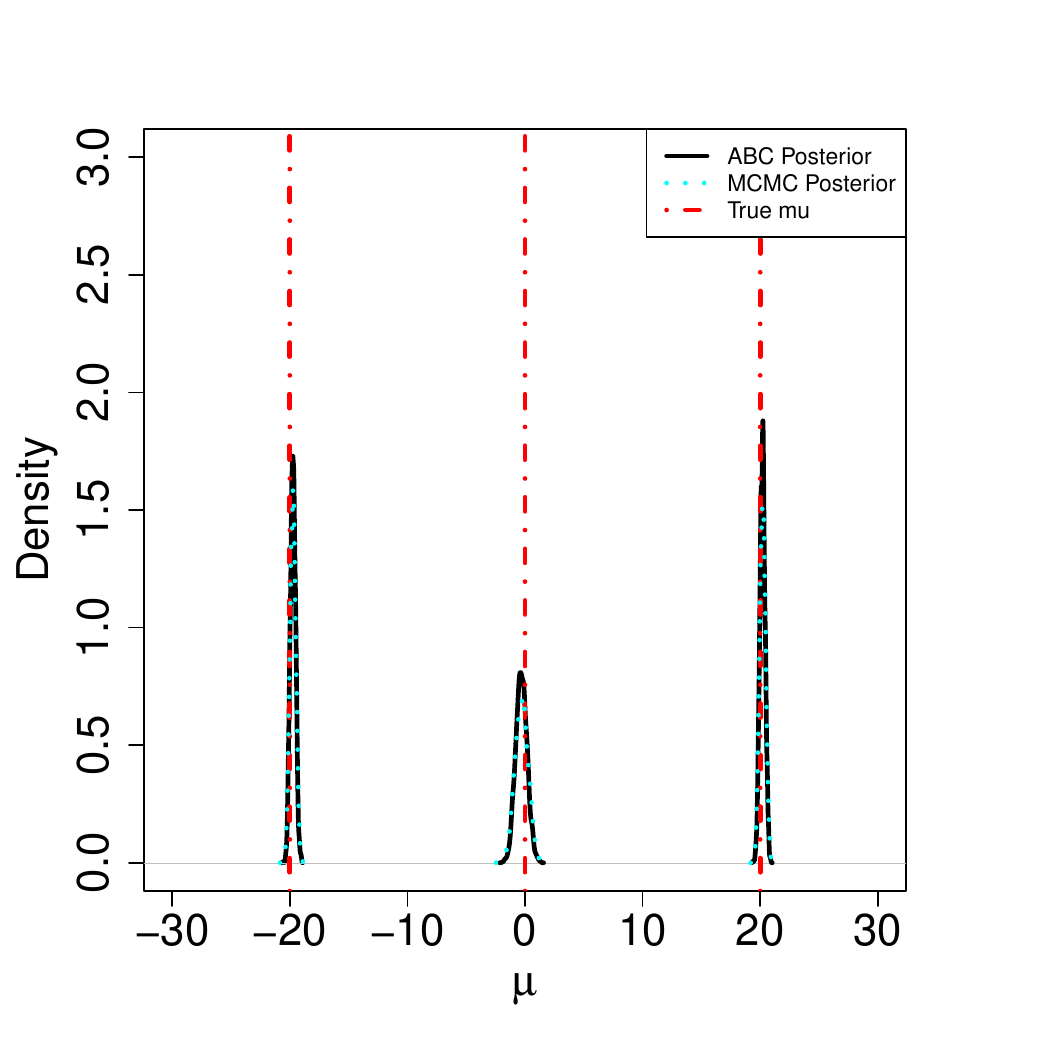}
    \caption{Marginal posteriors for $\mu$}\label{fig:ex1bis.mean}
    \end{subfigure}
    \qquad
    \begin{subfigure}{6cm}
    \includegraphics[height = 2.5in]{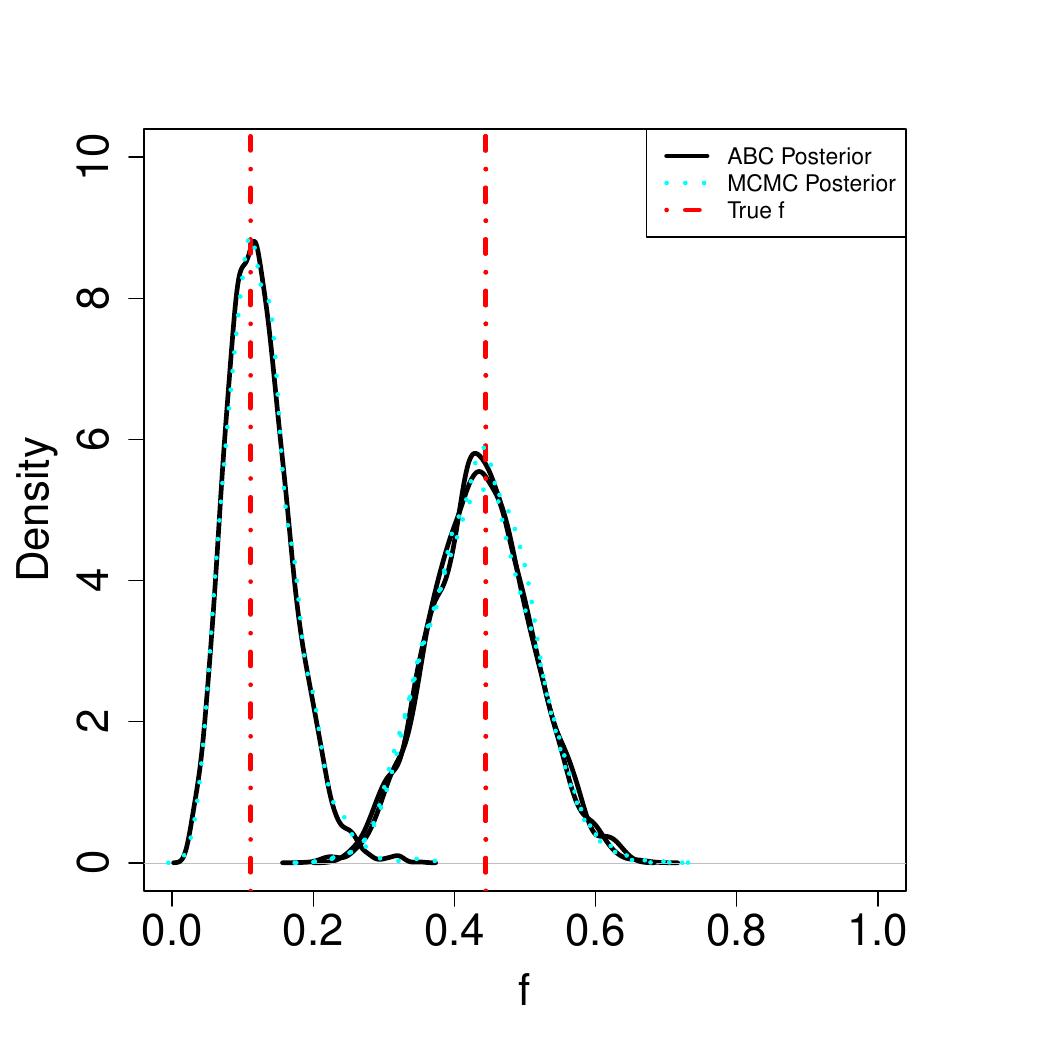}
    \caption{Marginal posteriors for $f$}\label{fig:ex1bis.f}
    \end{subfigure}
    \caption{Comparison between the ABC and the MCMC marginal posterior distributions for the three-component GMM example from Mena and Walker \cite{Ramses2015}. The true values for the means and the mixture weights are displayed as vertical dot-dashed red lines. 
   The final ABC posteriors obtained using the label switching (LS) procedure proposed in \S\ref{subsec.ls} are the solid black lines (ABC Posterior) while the MCMC posteriors are displayed with dotted cyan lines (MCMC Posterior).
   The number of particles for the ABC analysis and the number of elements kept from the MCMC analysis (after the burn-in) are equal to $5000$. 
   }%
    \label{Fig:example1bis}%
\end{figure}

Figure \ref{Fig:example1bis} shows the MCMC and the ABC posteriors for the weights and the means of the mixture components. 
The behavior of the ABC posterior distributions is consistent with their MCMC benchmarks. 
A summary of the results presented in Table~\ref{Table:example1bis} shows that the posterior means and the posterior standard deviations for the ABC posterior distributions are consistent with the ones retrieved using MCMC. Finally, in the third column, the Hellinger distances between the ABC and MCMC posteriors are provided. 
\begin{center}
\begin{table}
\centering
\begin{tabular}{|c|c|c|c|}
\hline
Parameter (input)    & MCMC (SD)          &   ABC (SD)  & H     \\
\hline
$f_{1} (0.44)$            & $0.44 (0.071)$    &  $0.44 (0.071)$   &  $0.024$     \\
\hline
$f_{2} (0.12)$          & $0.12 (0.048)$     &   $0.12 (0.048)$    &   $0.018$     \\
\hline
$f_{3} (0.44)$          & $0.44 (0.071)$      &  $0.44 (0.071)$     &   $0.033$    \\
\hline
$\mu_{1} (-20)$            & $ -19.61 (0.26) $     &  $-19.73 (0.22)$     &  $0.27$    \\
\hline
$\mu_{2} (0)$         & $ -0.33 (0.45)$     &  $ -0.30 (0.48) $    &    $0.17$     \\
\hline
$\mu_{3} (20)$         & $20.06 (0.24)$      &  $20.19 (0.22) $     &  $0.29$     \\
\hline
\end{tabular}
\caption{Posterior means (posterior standard deviations) obtained by using the MCMC and the ABC--PMC algorithm for the three-component GMM example from Mena and Walker \cite{Ramses2015}. 
The fourth column is the Hellinger distance between the final ABC posterior distribution and the MCMC posterior. The number of particles and the number of elements retained from the MCMC chain (after the burn-in) are both equal to $5000$.}
\label{Table:example1bis}
\end{table}
\end{center}

\subsection{Mixture Model with unequal group size} \label{sec.example2}
\begin{figure}[!ht]
    \centering
    \begin{subfigure}{6cm}
    \includegraphics[height = 2.5in]{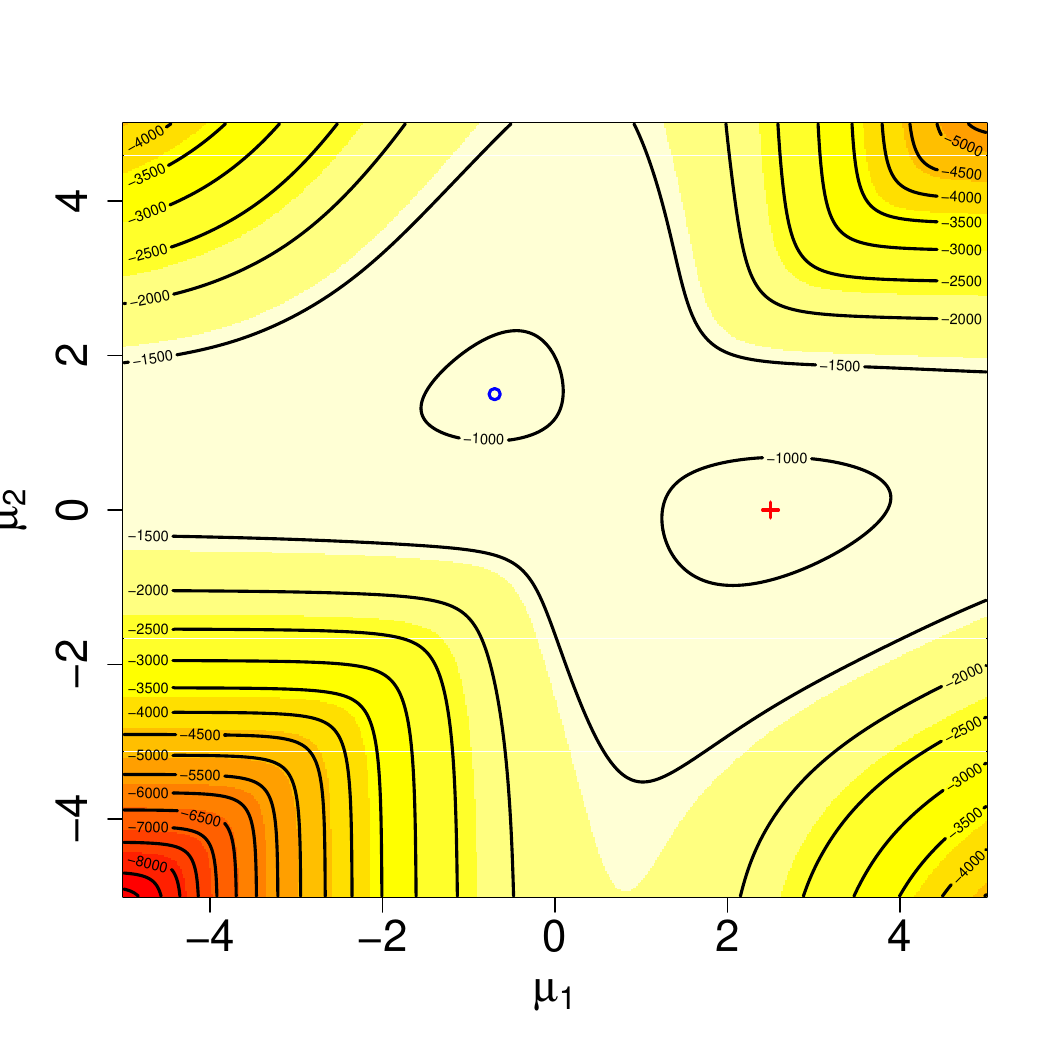}
    \caption{Log--likelihood surface for $\mu$}\label{fig:exMarin.LK}
    \end{subfigure}
    \qquad
    \begin{subfigure}{6cm}
    \includegraphics[height = 2.5in]{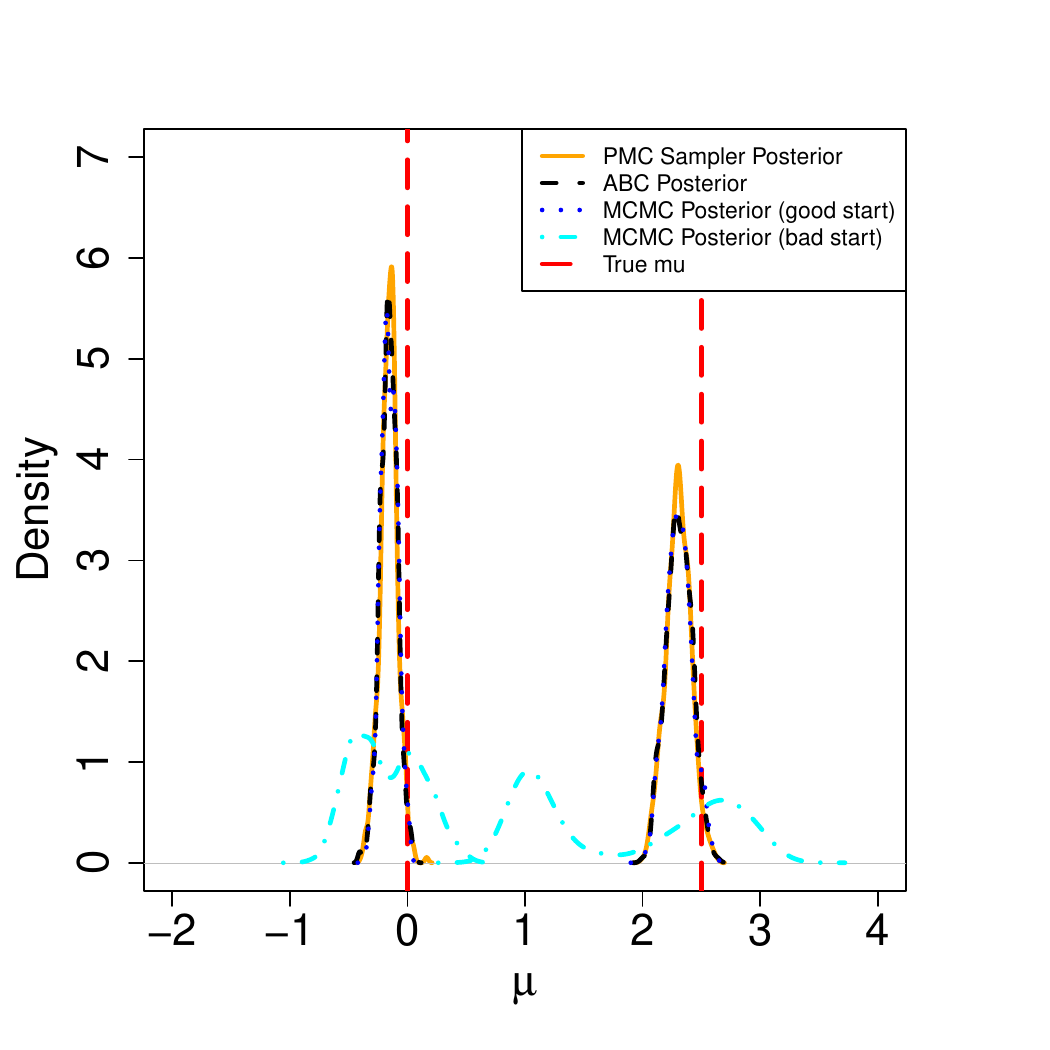}
    \caption{Marginal posteriors for $\mu$}\label{fig:exMarin.mu}
    \end{subfigure}
    \caption{(a) The log-likelihood surface of the Gaussian mixture model proposed by Marin et al. \cite{Marin2005}. There are two modes in the log-likelihood function, one close to the true value, ($2.5, 0$), highlighted with a red cross, and a local mode indicated with a blue circle.
   (b) The marginal PMC (solid orange lines), ABC (dashed black lines) and MCMC posterior distributions; the displayed MCMC posteriors include the results for good initial starting values as dotted blue lines (MCMC Posterior (good start)) and bad initial starting values as dot-dashed cyan lines (MCMC Posterior (bad start)). The true means are displayed as vertical long-dashed red lines.
   }%
    \label{figure:mixtureMarin}%
\end{figure}

%
Even in those cases in which the definition of the mixture model does not lead to the label switching problem, a second category of issues related to the multimodality of the likelihood function is present. This behavior has been studied from both the frequentist and the Bayesian perspectives. In particular, Marin et al. \cite{Marin2005} defined the following simple two-component mixture model to illustrate the multimodality issue,
\begin{equation}
f_1 \cdot \phi(\mu \mid \mu_{1},1)+(1-f_1) \cdot \phi(\mu \mid \mu_{2},1),
\label{eq:model1Marin}
\end{equation}
where the weight $f_1$ is assumed known and different from $0.5$ (avoiding the label switching problem). 
According to the specifications by  Marin et al. \cite{Marin2005}, $n=500$ samples were drawn from Eq.~\eqref{eq:model1Marin}, with $\theta=(f_1,\mu_{1},\mu_{2})=(0.7,0,2.5)$. 
The bimodality of the likelihood function (see Figure~\ref{figure:mixtureMarin}) makes the use of both the EM algorithm \citep{Dempster1977} and the Gibbs Sampler \citep{Diebolt1994} risky because their success depends on the set of initial values selected for initiating the algorithms.

The PMC sampler \citep{CappeEtAl2004, Marin2005} is used as a benchmark for the proposed ABC--PMC solution. 
Figure \ref{figure:mixtureMarin} displays the log likelihood function (note the two modes), and the final ABC--PMC posteriors with MCMC posteriors using good and bad starting values along with the posteriors using the PMC algorithm.
%
Table \ref{Table:example2} lists the means for the final ABC, MCMC, and PMC posteriors for $\mu_{1}$ and $\mu_{2}$, along with the Hellinger distance between the final ABC--PMC posteriors and the PMC posterior.  The ABC--PMC posteriors closely match the PMC posteriors.
\begin{center}
\begin{table}
\centering
\scalebox{1}{
\begin{tabular}{|c|c|c|}
\hline
Parameter (input)        &   $\mu_{1} (0)$   &   $\mu_{2} (2.5)$     \\
\hline
PMC (SD) &  $-0.17 (0.11)$ & $2.29 (0.17)$ \\
\hline
ABC--PMC (SD) &  $-0.16 (0.11)$  & $2.29 (0.17)$ \\
\hline
 MCMC$_{\text{good}}$ (SD) &   $-0.16 (0.11)$ & $2.31 (0.16)$    \\
\hline
 MCMC$_{\text{bad}}$ (SD) &    $1.86 (0.80)$ & $-0.18 (0.29)$    \\
 \hline
 H &  $0.051$  &  $0.048$   \\
\hline
\end{tabular}}
\caption{Posterior means (posterior standard deviations) obtained by using MCMC (with good and poor choices for initializing the procedure), PMC, and ABC--PMC algorithms in the example by Marin et al. \cite{Marin2005}. The last column indicates the Hellinger distance between the final ABC posterior distributions and the PMC posteriors}
\label{Table:example2}
\end{table}
\end{center}

\section{Application to Galaxy Data} \label{sec.galaxy}
The galaxy dataset was introduced to the statistical community in Roeder \cite{Roeder1990}, and has been commonly used for testing clustering methods. The data contains the recessional velocities of $82$ galaxies (km/sec) from six well separated sections of the Corona Borealis region. The galaxy data has been well-studied by the statistical community \citep{Richardson1997, Roeder1997, Lau2007, Wang2012}.  
The recessional velocities of the galaxies are typically considered realizations of independent and identically distributed Gaussian random variables, but there is discrepancy in the conclusions about the number of groups in the GMM; estimates vary from three components \citep{Roeder1997} to five or six \citep{Richardson1997}.

In this analysis, we focused on the model with three components \citep{Roeder1997}, in order to be consistent with \cite{Ramses2015} and \cite{Marin2005}. 
For the hyperparameters, we use the Empirical Bayes approach suggested by Casella et al. \cite{Casella2000}. 
Additionally, since each recessional velocity was also assigned a measurement error, the ABC forward model has been modified to take into account this information. 
In order to include the measurement errors in the forward model, each simulated recessional velocity is assigned one of the observed measurement errors. The simulated and observed recessional velocities were matched according to their ranks, and the measurement error of the observations were assigned to the simulated data according to this matching. Then, Gaussian noise was added to each simulated recessional velocity with a standard deviation equal to its assigned measurement error.

The posterior means for each component's parameters are listed in Table \ref{Table:example3}.  
The third component was found to have a weight equal to $0.057$, and a mean and variance equal to $32.94$ $km/sec$ and $1.16$ $km^2/sec^2$, respectively. 
The main difference with the proposed ABC--PMC estimates and those from Marin et al. \cite{Marin2005}, in which they also fixed $K = 3$, is with the estimated mean and variance of the third component; however, the proposed ABC--PMC estimates are consistent with those obtained by Mena and Walker \cite{Ramses2015}. 

\begin{figure}[htbp]
   \centering
\includegraphics[height = 2.3in]{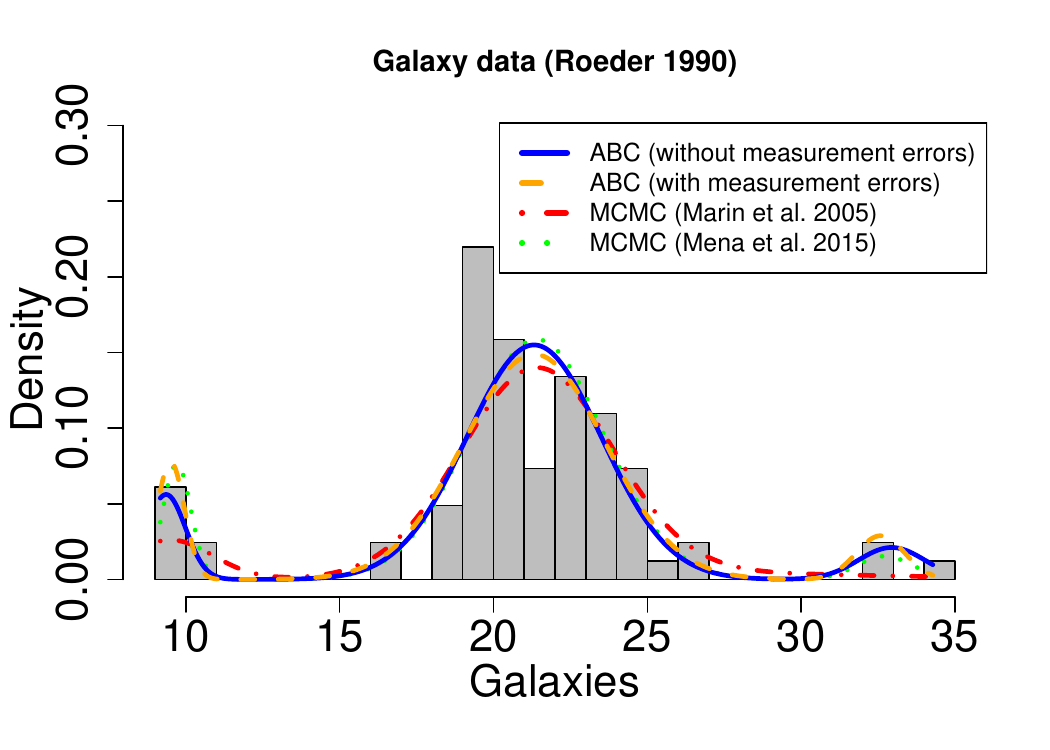} 
\caption{Histogram of the recessional velocity of 82 galaxies and the estimated three-component Gaussian mixture models for each study. The results obtained by the proposed ABC--PMC algorithm are comparable with the ones obtained using MCMC. When accounting for the measurement error in the data, the first and the third component of the mixture model have more clearly defined modes (dashed orange line). The posterior means for the mixture weights, means and variances used are displayed in Table~\ref{Table:example3}.}
   \label{Fig:exampleGalaxy_comparisons}
\end{figure}

By using the additional information about the measurement errors, the proposed ABC--PMC algorithm can provide posterior distributions that better reflect the uncertainty in the observations.
Including measurement errors in the forward model affects the resulting ABC posterior as reported in Table \ref{Table:example3} and the relative estimates plotted in orange in Figure \ref{Fig:exampleGalaxy_comparisons}. In particular, the estimated variances are smaller for the proposed ABC--PMC algorithm when measurement errors are accounted for, which results in more clearly defined clusters for the first and third components of the mixture model.

\begin{center}
\begin{table}
\centering
\scalebox{0.7}{
\begin{tabular}{|c|c|c|c|c|}
\hline
Parameter          & Marin et al. (2005)        & Mena et al. (2015) &   ABC--PMC & ABC--PMC\tiny{(with errors)}   \\
\hline
$f_{1}$            & 0.09    &  0.087    &   0.089  & 0.087 \\
\hline
$f_{2}$          & 0.85     &    0.868     & 0.85 & 0.86 \\
\hline
$f_{3}$          & 0.06     &     0.035   &   0.061 & 0.053\\
\hline
$\mu_{1} (km/sec)$            & 9.5     &   9.71   &  9.36  & 9.51   \\
\hline
$\mu_{2} (km/sec)$         & 21.4    &     21.4    &  21.32 & 21.33 \\
\hline
$\mu_{3} (km/sec)$         & 26.8     &    32.72     &  32.94  & 32.58\\
\hline
$\sigma_{1}^{2} (km^2/sec^2)$            & 1.9    &    0.21    & 0.40  & 0.20   \\
\hline
$\sigma_{2}^{2} (km^2/sec^2)$         &  6.1    &     4.76    & 5.32 &   4.79 \\
\hline
$\sigma_{3}^{2} (km^2/sec^2)$         & 34.1     &     0.82     & 1.16  & 0.62 \\
\hline
\end{tabular}}
\caption{Comparison between the posterior means obtained by Marin et al. \cite{Marin2005}, Mena and Walker \cite{Ramses2015} (MCMC algorithm) and the proposed ABC--PMC algorithm for the Galaxy data. The results of the ABC--PMC analysis including measurement errors are displayed in the fourth column.}
\label{Table:example3}
\end{table}
\end{center}

\section{Concluding Remarks} \label{sec.conclusions}

The popularity of ABC is, at least in part, due to its capacity to handle complex models.  Extensions of the basic algorithm have lead to improved efficiency of the sampling, such as 
the ABC--PMC algorithm of Beaumont et al. \cite{BeaumontEtAl2009}.  We propose an ABC--PMC algorithm that can successfully handle finite mixture models.
Some of the challenges with inference for finite mixture models are due to the complexity of the likelihood function including its possible multimodality and the exchangeability of the mixture component labels leading to the label switching problem.
Fortunately, ABC can handle complicated likelihood functions, but the label switching problem needs to be addressed.  
We suggest a procedure for addressing the label switching problem within the proposed ABC algorithm that works well empirically.
Some additional challenges with using ABC for mixture models include the selection of informative summary statistics, and defining a kernel for moving the mixture weights since they are constrained to be between 0 and 1 and must sum to 1.
For the summary statistics, we propose using the Hellinger distance between kernel density estimates of the real and simulated observations, which allows the multimodality of the data to be accounted for and compared between the two sets of data.
We propose a Dirichlet resampling algorithm for moving the mixture component weights that preserves some information from the sampled particle, but also improves the efficiency of the ABC--PMC procedure (by not having to draw from the Dirichlet prior at each time step). 

The proposed ABC algorithm has been explored and tested empirically using popular examples from the literature.  The resulting ABC posteriors were compared to the corresponding MCMC posteriors, and in all cases considered the proposed ABC and MCMC posteriors were similar, as desired.  
We also considered the galaxy velocity data from the Corona Borealis Region \citep{Roeder1990}, which is often used in assessing the performance of mixture models.  
An advantage of ABC over other commonly used methods is that the forward model can be easily expanded to better represent the physical process that is being modeled.  For the galaxy data, measurement errors are available in the original data set, but are not generally used when analyzing the data.  We extend the proposed ABC--PMC forward model used in this example to include the measurement errors, which provides a more accurate assessment of the uncertainty in the data.

Though the presented examples focused on one--dimensional GMMs, the component distributions can be adapted to other distributions in the ABC forward model.  Overall, the proposed ABC--PMC algorithm performs well and is able to recover the benchmark MCMC posteriors (when they were available) suggesting that the proposed ABC--PMC algorithm is a viable approach for carrying out inference for FMMs.


\appendix

\section{Mixture weight resampling kernel} \label{appendix}
The steps for resampling the mixture weights at iteration $t$, $\{f_t^{(J)}\}_{J=1}^N$, are outlined in Algorithm~\ref{alg:dr} with details presented in Section~\ref{subsec.dirichlet.res}.  
In this Section, we present mathematical justification for the proposed resampling procedure.
In particular, we want to show that the resampled weights, $f_t = (f_{t,1}, \ldots, f_{t,K})$ with $f_{t,i} = \xi_{t,i}^*/\xi_{t,+}^*$, $i = 1, \ldots, K$, $\xi_{t,+}^*=\sum_{i = 1}^K\xi_{t,i}^*$, and $t = 2, \ldots, T$, follows a Dirichlet$(\delta_1, \ldots, \delta_K)$.

Recall that to resample a particle at iteration $t$, a particle from the previous iteration, $t-1$, is randomly sampled according to the importance weights, $\{W_{t-1}^{(J)}/\sum_{L = 1}^NW_{t-1}^{(L)}\}_{J=1}^N$.
Suppose some mixture weight vector $f_{t-1}^{(J*)} = f_{t-1} = \{f_{t-1,1}, \ldots, f_{t-1,K}\}$, $t \in \{2, \ldots, T\}$ has been selected for resampling with  some finite number of components $K>1$.  To show that carrying out the resampling steps as outlined in Algorithm~\ref{alg:dr} results in $f_t\sim$Dirichlet$(\delta_1, \ldots, \delta_K)$, consider the following propositions.
Proposition~\ref{prop1} and \ref{prop4} correspond to Step 5 of Algorithm~\ref{alg:dr}, and Proposition~\ref{prop2} and \ref{prop3} correspond to Steps 1 and 3, respectively.

\begin{prop} \label{prop1}
Let $f_{t-1} = \left(\frac{\xi_{t-1,1}}{\xi_{t-1,+}}, \ldots, \frac{\xi_{t-1,K}}{\xi_{t-1,+}}\right)$ where $\xi_{t-1,i} \myeq \text{Gamma}(\delta_i, \beta)$ and $\xi_{t-1,+} = \sum_{i = 1}^K \xi_{t-1,i}$, then $f_{t-1} \sim  \text{Dirichlet}(\delta)$.
\end{prop}

\begin{proof}
This is a well-known result that can be proved by using the change-of-variables approach with $y_1 = \xi_{t-1,1}/y_K, \ldots, y_{K-1}=\xi_{t-1,K-1}/y_K$, and $y_K = \xi_{t-1,+}$.  The determinant of the Jacobian of the transformation is $y_K^{K-1}$.  Noting the joint distribution $\xi_{t-1,1}, \ldots, \xi_{t-1,K} \sim  e^{-\sum_{i=1}^K \beta \xi_{t-1,i}}\prod_{i = 1}^K \beta
^{\delta_i} \frac{\xi_{t-1,i}^{\delta_i-1}}{\Gamma(\delta_i)}$, then the joint distribution using the change-of-variables is
$$
y_1, \ldots, y_K \sim \beta^{\sum_{i=1}^K\delta_i} e^{- \beta y_K} \left(y_K^{K-1}\right) \left(y_K \left(1 - \sum_{i = 1}^{K-1} y_i\right)\right)^{\delta_K-1}\frac{\prod_{i = 1}^{K-1}(y_iy_K)^{\delta_i-1}}{\prod_{i = 1}^{K}\Gamma(\delta_i)}.
$$
Then, integrating out the $y_K$ results in 
$$
y_1, \ldots, y_{K-1} \sim \left(1 - \sum_{i = 1}^{K-1} y_i\right)^{\delta_K-1}\frac{\Gamma(\sum_{i=1}^K\delta_i)}{\prod_{i = 1}^{K}\Gamma(\delta_i)}\prod_{i = 1}^{K-1}y_i^{\delta_i-1}.
$$
Noting that $0\leq y_1, \ldots, y_{K-1} \leq 1$ and $0 \leq \sum_{i=1}^{K-1} y_i \leq 1$, then we can let $y_K = 1 - \sum_{i = 1}^{K-1} y_i$, resulting in
$$
y_1, \ldots, y_K \sim \frac{\Gamma(\sum_{i=1}^K\delta_i)}{\prod_{i = 1}^{K}\Gamma(\delta_i)}\prod_{i = 1}^K y_i^{\delta_i-1}, 
$$
which is a Dirichlet($\delta_1, \ldots, \delta_K$) distribution.
\end{proof}

\begin{prop} \label{prop2}
Let $f_{t-1}  \sim \text{Dirichlet}(\delta)$ with $\delta = (\delta_1, \ldots, \delta_K)$ and $Z_t \sim \text{Gamma}(\delta_+, \beta)$ where $\delta_+ = \sum_{i = 1}^K \delta_i$.  Then $\xi_{t,i} = Z_tf_{t-1,i} \sim \text{Gamma}(\delta_i, \beta)$.
\end{prop}

\begin{proof}
This follows from Theorem 1 of Yeo and Milne \cite{Yeo:1991aa} by noting that the marginal distribution of a Dirichlet($\delta$), that is, the distribution of $f_{t-1,i}$, is a Beta($\delta_i, \sum_{i = 1}^K \delta_j - \delta_i)$.
\end{proof}

\begin{prop} \label{prop3}
Given $p \in  [0,1]$ and independent $B_{t,i} \sim \text{Beta}(p\delta_i, (1-p)\delta_i)$ for $i = 1, \ldots, K$, then $\xi_{t,i}B_{t,i} \sim \text{Gamma}(p\delta_i, \beta)$.
\end{prop}

\begin{proof}
This follows directly from Theorem 1 of Yeo and Milne \cite{Yeo:1991aa}.
\end{proof}

\begin{prop} \label{prop4}
Given $p \in  [0,1]$,  independent $B_{t,i} \sim \text{Beta}(p\delta_i, (1-p)\delta_i)$, independent $\eta_{t,i} \myeq \text{Gamma}((1-p)\delta_{i}, \beta)$, $i = 1, \ldots, K$, 
and set $\xi^{*}_{t,i} = Z_tf_{t-1,i}B_{t,i} + \eta_{t,i}$ and $f_{t,i} = \xi_{t,i}^*/\xi_{t,+}^*$, with $\xi_{t,+}^*=\sum_{i=1}^{K} \xi_{t,i}^*$.  Then $f_t = (f_{t,1}, \ldots, f_{t,K}) \sim \text{Dirichlet}(\delta_1, \ldots, \delta_K)$.
\end{prop}

\begin{proof}
By Propositions~\ref{prop2} and \ref{prop3}, $Z_tf_{t-1,i}B_{t,i} \sim \text{Gamma}(p\delta_i, \beta)$, and then $\xi^{*}_{t,i}\sim\text{Gamma}(\delta_i, \beta)$ by noting that the sum of the two independent gamma random variables with the same rate parameter, $\beta$, is also a gamma random variable with rate parameter $\beta$ and with the sum $\delta_i = p\delta_1+(1-p)\delta_1$ as the shape parameter.  Then Proposition~\ref{prop1} gives the distribution for $f_t$.
\end{proof}

\bibliographystyle{unsrt}
\bibliography{mybib.bib}

\end{document}